\documentclass[aps,twocolumn,showpacs,prd,aps,floatfix,superscriptaddress]{revtex4-1}
\usepackage{bm,multirow,amssymb,amsbsy,amsmath,graphicx,bbold,color}
\makeatletter
\usepackage{pifont,bbm}
\makeatother

\newcommand{\mathd}{\mathrm{d}}
\usepackage{color}
\newcommand{\ket}[1]{\vert #1 \rangle}
\newcommand{\bra}[1]{\langle #1 \vert}
\newcommand{\braket}[2]{\langle #1 | #2 \rangle}
\newcommand{\rc}{r_C}
\newcommand{\mom}{Q}
\newcommand{\scalar}[2]{\langle  #1 \vert #2 \rangle }
\begin{document}
\title{Dissipative Continuous Spontaneous Localization (CSL) model}
\author{Andrea \surname{Smirne}}
\email{andrea.smirne@ts.infn.it}
\affiliation{Dipartimento di Fisica, Universit{\`a} degli Studi di Trieste, Strada Costiera 11, I-34151 Trieste, Italy}
\affiliation{Istituto Nazionale di Fisica Nucleare, Sezione di Trieste, Via Valerio 2, I-34127 Trieste, Italy}
\author{Angelo \surname{Bassi}}
\email{bassi@ts.infn.it}
\affiliation{Dipartimento di Fisica, Universit{\`a} degli Studi di Trieste, Strada Costiera 11, I-34151 Trieste, Italy}
\affiliation{Istituto Nazionale di Fisica Nucleare, Sezione di Trieste, Via Valerio 2, I-34127 Trieste, Italy}
\begin{abstract}
Collapse models explain the absence of quantum superpositions at the macroscopic scale,
while giving practically the same predictions as quantum mechanics for microscopic systems. The Continuous
Spontaneous Localization (CSL) model is the most refined and studied among collapse models.
A well-known problem of this model, and of similar ones, is the steady and unlimited increase of the energy induced by the collapse noise.
Here we present the dissipative version of the CSL model, which
guarantees a finite energy during
the entire system's evolution, thus making a crucial step toward a realistic
energy-conserving collapse model. This is achieved by introducing
a non-linear stochastic modification of the Schr{\"o}dinger equation,
which represents the action of a dissipative finite-temperature collapse noise.
The possibility to introduce dissipation within collapse models in a consistent way
will have relevant impact on the experimental investigations of the CSL model, 
and therefore also on the testability of the quantum superposition principle.
\end{abstract}

\pacs{03.65.Ta, 03.65.Yz, 05.40.-a} 
\maketitle
The superposition principle lies at the core of quantum mechanics.
The last years have experienced a huge progress in the theoretical
and experimental 
investigation aimed at preparing and observing quantum superpositions
of large systems \cite{Friedman2000,Zawisky2002,Gerlich2007,Dimopoulos2009,Romero-Isart2011,Nimmrichter2011,Arndt2014}.
Such a progress promises a crucial insight into a question
which was born with quantum mechanics itself \cite{Bohr1928}.
Can quantum mechanics be applied at all scales, including
the macroscopic ones, or is there an intrinsic limit,
above which its description of reality is not appropriate?
Collapse models \cite{Ghirardi1986,Diosi1989,Ghirardi1990,Bassi2003,Bassi2013}
show explicitly how the second point of view can be assumed without the need 
to introduce
an ad-hoc separation between the
microscopic and the macroscopic world within the theory~\cite{Bell1987}.
Through a non-linear stochastic modification of the Schr{\"o}dinger equation, 
collapse models predict a behavior of microscopic systems which follows almost strictly
that of standard quantum mechanics, while
preventing macroscopic systems from being
in a superposition of macroscopically distinct positions.

The continuous spontaneous localization (CSL) model \cite{Ghirardi1990} is
the most refined collapse model, as it also applies to identical particles.
The mass density of a quantum system is coupled with a white-noise
field, which can be interpreted as a classical random field filling space \cite{Bassi2013}.
Different speculations on the origin of the noise field have been developed,
tracing it back, e.g., to gravity \cite{Diosi1987} or to cosmological particles~\cite{Adler2008}.
However, the full characterization of such a noise 
calls for a new fundamental theory, which departs from quantum mechanics
and can explain the classical nature of the noise, as well as its non-hermitian and non-linear coupling
with matter \cite{Bassi2013,Adler2014}. 
In this respect, the CSL model, and every collapse model, should be seen as
a phenomenological model expressing the influence
of the noise field in an effective way.

The localization of the wavefunction of macroscopic objects, along with the
resulting destruction of quantum superpositions, is not the only distinctive
feature of the CSL model with respect to the usual Schr{\"o}dinger evolution.
The action of the noise induces a steady increase of the mean kinetic energy,
which diverges on the long time scale \cite{Ghirardi1990}, thus
manifestly leading to a violation of the principle of energy conservation.
Despite the smallness of the increase rate,
the comparison of the predictions on 
the secular energy with cosmological data
provides some of the strongest experimental bounds on the two intrinsic parameters of the model \cite{Adler2007,Adler2009}.
In particular, the spontaneous heating of the intergalactic medium which would be induced by the stochastic noise
sets  $\lambda \sim 10^{-9}\text{s}^{-1}$ as an upper bound to the localization rate $\lambda$. 
This value coincides with the proposal by Adler based on latent image formation in photography \cite{Adler2007}.

A significant and long-time debated  \cite{Ballentine1991,Fleming1991,Pearle1996,Bassi2005,Vacchini2007}
issue is whether the
divergence of the energy in collapse models can be avoided,
thus pointing to a reestablishment of the energy conservation principle,
while preserving the specific features any collapse model must have. 
In this work, we prove that this can be attained by properly extending the CSL model. 
We modify the defining stochastic differential equation
via the introduction of new operators,
which depend on the momentum of the system.
This determines the occurrence of
dissipation \cite{Vacchini2000,Hornberger2006,Vacchini2009}, thus leading
to the relaxation of the energy to a finite asymptotic value.
The latter can be associated with a finite temperature of the noise field.
Remarkably, such a temperature can take on small values
(of the order of 1 K) while the effectiveness of the model is maintained.
Contrary to a common misconception, the steady increase of the energy is not an
unavoidable trait of the collapse models inducing localization in space: in our dissipative model 
there is a continuous localization of the wavefunction, while the mean energy of the system will typically decrease. 

The paper is organized as follows. In Sec. \ref{sec:tcm}, we briefly recall the main features of the 
original CSL model put forward in \cite{Ghirardi1990}. In Sec. \ref{sec:deo}, we introduce the dissipative extension
of the CSL model via a proper stochastic differential equation, which requires the definition of a new parameter
within the model.
The defining operators are interpreted by moving to the momentum representation,
in analogy with an external momentum-dependent potential.
In Sec. \ref{sec:era}, after deriving the master equation for the statistical
operator associated with the model, we prove explicitly that the mean kinetic energy
relaxes to a finite value, whose natural interpretation is given
in terms of a temperature associated with the noise field.
Crucially, even in the presence of a low temperature noise one can have an effective
collapse model, which leads to reconsider the bounds on the CSL parameters due to the long-time
energy behavior. In Sec. \ref{sec:mol}, we show that our
extended model still provides a unified description of microscopic and macroscopic systems by virtue of the so-called amplification mechanism;
the relation between the time scales of energy relaxation and position localization is also discussed.
Final remarks and possible future perspectives are given in Sec. \ref{sec:c}.

\section{The CSL model}\label{sec:tcm}
Using the language of non-relativistic quantum field theory,
the CSL model is formulated in terms of a stochastic differential equation in the Fock space
associated with the system \cite{Ghirardi1990}.
Given different types of particles, where the type $j$ has mass $m_j$,
the mass-proportional CSL model \cite{Pearle1994} is defined by
\begin{eqnarray}\label{eq:sdecsl}
\mathd \ket{\varphi_t} &=& \left[-\frac{i}{\hbar}\hat{H} \mathd t + \frac{\sqrt{\gamma}}{m_0} \int \mathd {\bf y} [\hat{M}({\bf y})-\langle M({\bf y}) \rangle_t ] \mathd W_t({\bf y}) \right. \nonumber \\
&&\left.- \frac{\gamma}{2 m^2_0} \int \mathd{\bf y}  [\hat{M}({\bf y})-\langle M({\bf y}) \rangle_t ]^2 \mathd t \right] \ket{\varphi_t},
\end{eqnarray}
where $\hat{H}$ is the standard quantum Hamiltonian, $\langle A \rangle_t \equiv \bra{\varphi_t} \hat{A} \ket{\varphi_t}$, $m_0$ is a reference mass (usually the mass of a nucleon)
and $\hat{M}({\bf y})$ is a smeared mass density operator:
\begin{equation}
\hat{M}({\bf y}) = \sum_j m_j \int \frac{\mathd {\bf x}}{(\sqrt{2 \pi} r_C)^3} e^{-\frac{|{\bf y}-{\bf x}|^2}{2 r^2_C}} \hat{\psi}_j^{\dag}({\bf x}) \hat{\psi}_j({\bf x}).
\end{equation}
Here, $\hat{\psi}_j^{\dag}({\bf x})$ and $\hat{\psi}_j({\bf x})$ are, respectively, the creation and the annihilation 
operator of a particle of type $j$ in the point ${\bf x}$,
while $W_t({\bf y})$ is an ensemble of independent Wiener processes,
one for each point in space. The model
is characterized by two parameters: $\gamma$,
which sets the strength of the collapse process, and $r_C$, 
which determines the threshold above which spatial superpositions are suppressed.
The choice of the numerical values for these parameters is of course ultimately dictated by the agreement
with the experimental data; the originally proposed values are \cite{Ghirardi1990} $r_C = 10^{-7} \text{m}$ and $\gamma = 10^{-30} \text{cm}^{3}\text{s}^{-1}$.

\begin{figure*}[!ht]
\hskip-3cm{\bf (a)}\hskip5cm{\bf (b)}\hskip6cm{\bf (c)}\\
\includegraphics[width=.65\columnwidth]{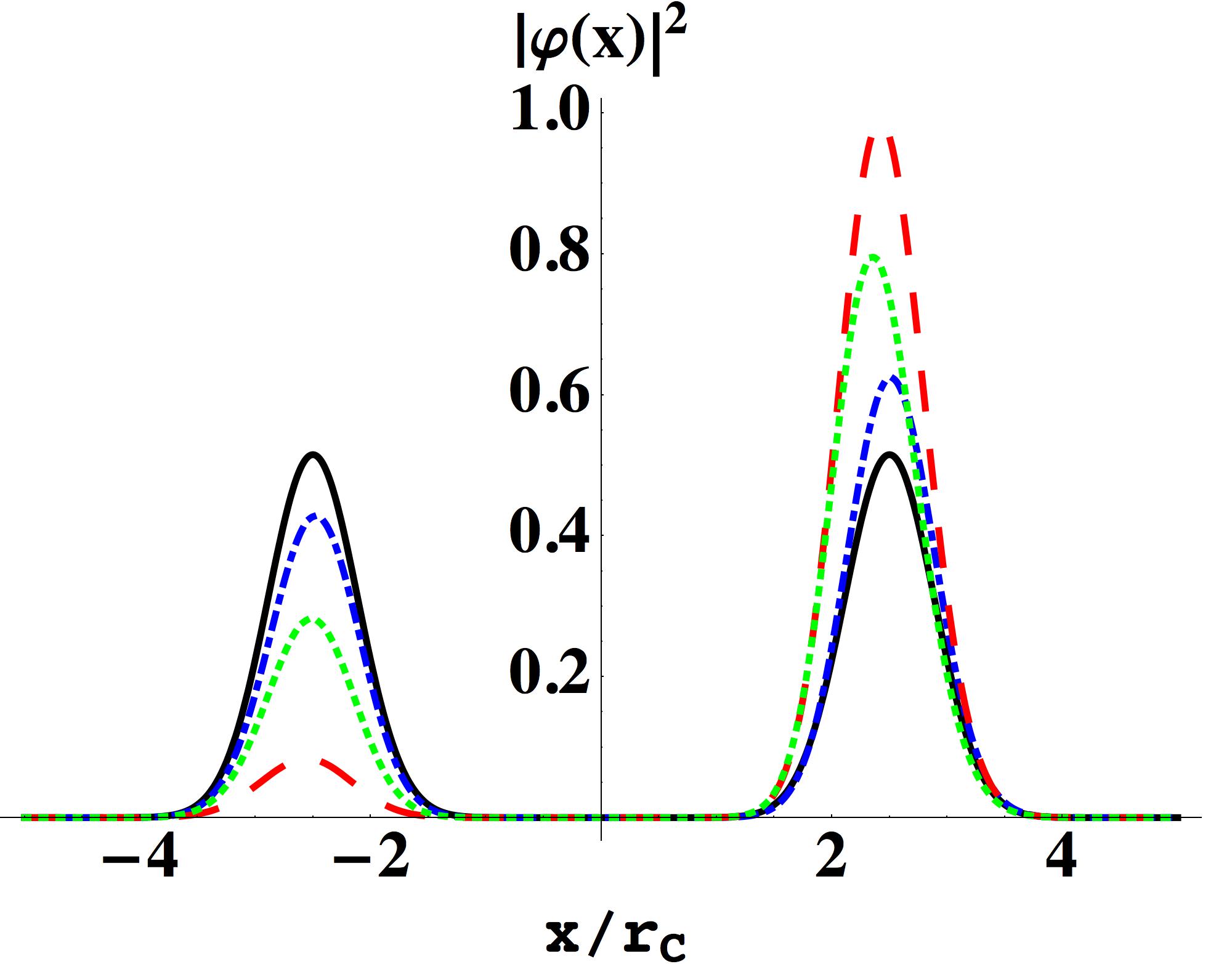}\hspace{0.3cm}\includegraphics[width=.65\columnwidth]{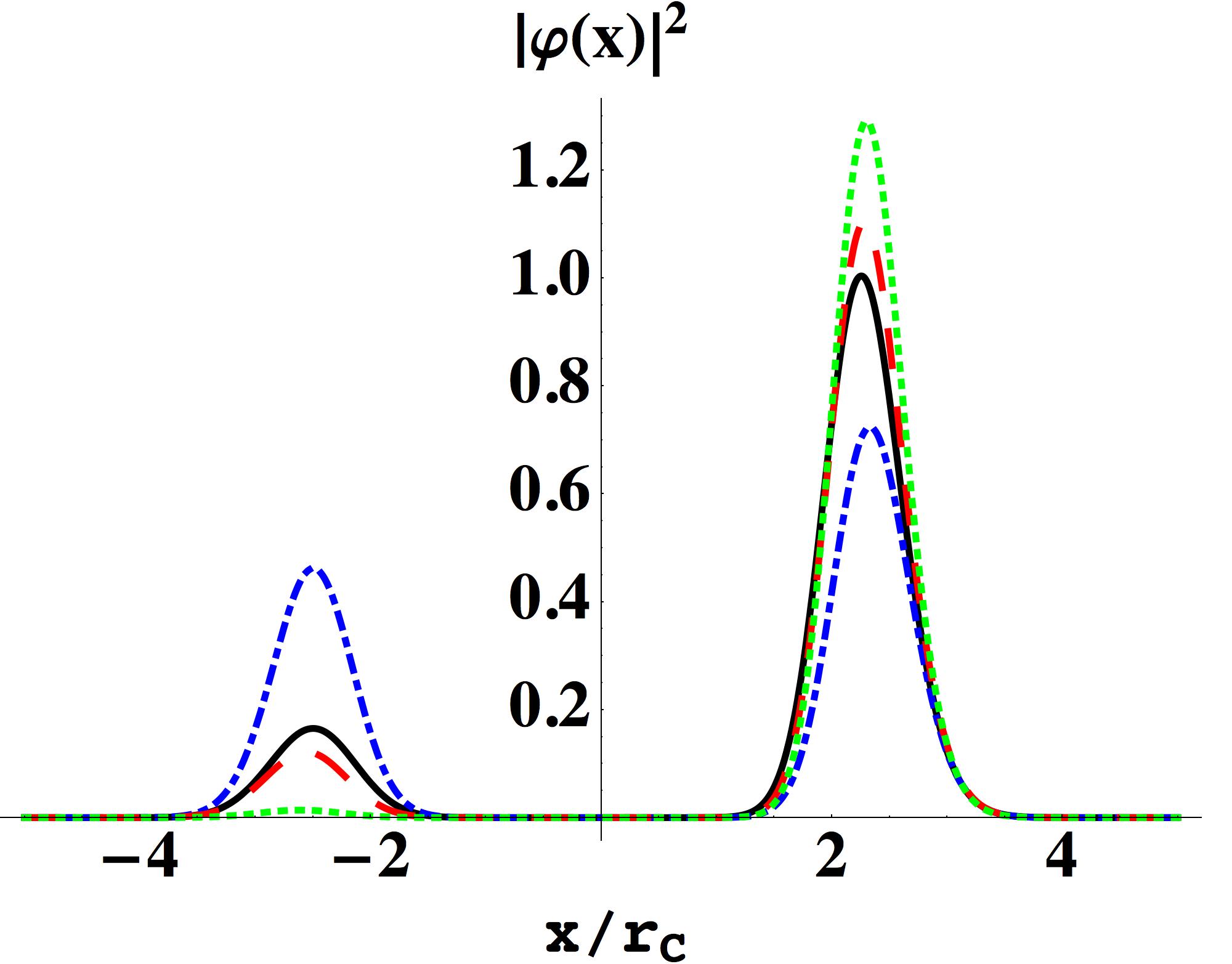}\hspace{0.3cm}\includegraphics[width=.65\columnwidth]{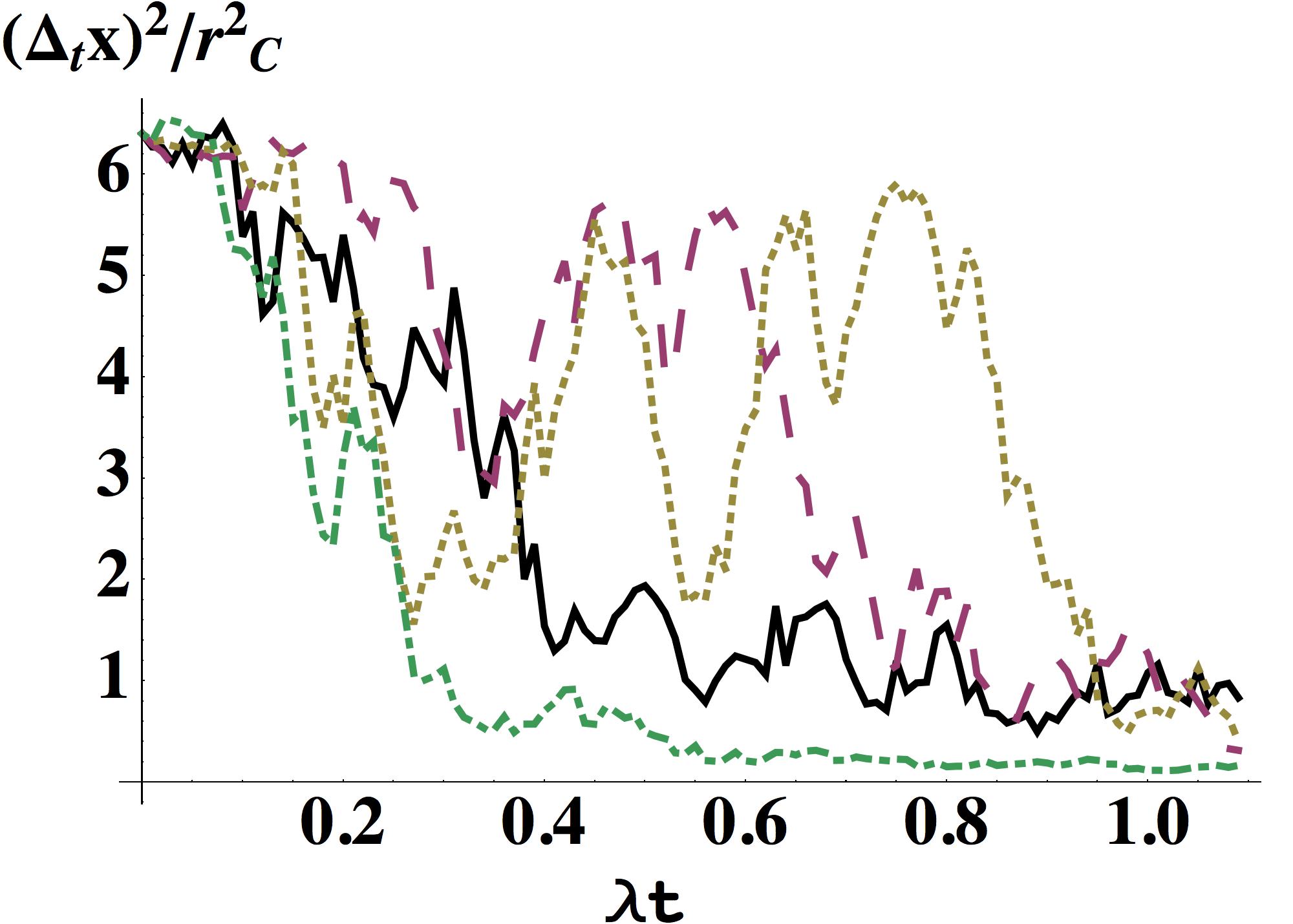}
\caption{{\bf (a)} and {\bf (b)} Evolution of the position probability distribution $|\varphi_t(x)|^2 = |\braket{x}{\varphi_t}|^2$ in the CSL model in one dimension, for one nucleon
initially in a balanced
superposition of two gaussian states with equal variance $\sigma^2$ and centered, respectively, in $\alpha$ and $-\alpha$.
The probability distribution is plotted for a single realization of the random noise and at times
$\lambda t = 0$ (black solid line), $\lambda t = 0.1$ (blue dot-dashed line),
$\lambda t = 0.3$ (red dashed line) and $\lambda t = 0.4$ (green dotted line) in {\bf (a)}, while $\lambda t = 0.5$ (black solid line), $\lambda t = 0.6$ (blue dot-dashed line),
$\lambda t = 0.8$ (red dashed line) and $\lambda t = 0.9$ (green dotted line) in {\bf (b)}; $\sigma/r_C = 0.55$ and $\alpha/r_C=2.5$. 
{\bf (c)} Time evolution of the position variance, $(\Delta_t x)^2 = \bra{\varphi_t} \hat{x}^2 \ket{\varphi_t}-(\bra{\varphi_t} \hat{x} \ket{\varphi_t})^2$, for different realizations of the noise field.
We have applied the Euler-Maruyama method \cite{Kloeden1992,Semina2014} to Eq.(\ref{eq:sdecsl}), for $\hat{H}=0$ and time step $\lambda \Delta t = 0.01$.
As discussed in the text, the rate $\lambda$ has to be replaced by the rate $\Gamma$ defined in Eq.(\ref{eq:G})
if a macroscopic object is taken into account, in accordance with the amplification mechanism.}
\label{fig:1}
\end{figure*}

The mass density operators $\hat{M}(\bf y)$ in Eq.(\ref{eq:sdecsl})
induce
a collapse of the wavefunction $\ket{\varphi_t}$
around the common eigenvectors of the position operators of the particles
composing the system \cite{Ghirardi1990}.
Hence, the asymptotic wavefunction is sharply localized around 
defined positions, excluding possible spatial superpositions.
The collapse rate for a microscopic system is given by $\lambda = \gamma/(4 \pi r^2_C)^{3/2} \approx 2.2 \times 10^{-17} \text{s}^{-1}$.
Such a small value guarantees that the spatial localization
due to the noise field can be safely neglected if a microscopic system
is taken into account. 
Now instead, consider a macroscopic rigid body in a superposition
of two states distant more than $r_C$.
Its center of mass collapses with an effective rate \cite{Adler2007,Bassi2014}
\begin{equation}\label{eq:G}
\Gamma = \lambda n^2 \tilde{N},
\end{equation}
where $n$ is the number of constituents of the body
contained in a volume $r_C^3$ and $\tilde{N}$
denotes how many such volumes are held in the macroscopic body.  
This relation clearly shows the amplification mechanism, which is at the basis
of every collapse model. The localization induced by the noise field
grows with the size of the system, so that the center of mass
of any macroscopic object behaves, for all practical purposes, according to classical mechanics.
The peculiar property of the CSL model is the quadratic dependence of the rate $\Gamma$
on the number of constituents, which is a direct
consequence of the action of the noise field on identical particles \cite{Bassi2013}.
The main features of the CSL model are summarized in Fig.\ref{fig:1},
where we represent the time evolution of the position probability distribution of one particle, which is
initially in a superposition of two gaussian states. 
The wavefunction is subjected continuously to the action of the noise, which suppresses the superposition between the two gaussians, leading to a gaussian state
localized around one of the two initial peaks, in a time scale fixed by the collapse rate, see Fig.\ref{fig:1} {\bf (a)} and {\bf (b)}. 
The diffusive nature of the dynamics in the CSL model is clearly illustrated by the
time-evolution of the position variance, see Fig.\ref{fig:1} {\bf (c)}.

A relevant drawback of the original CSL model,
as well as of most collapse models, is that the average kinetic
energy of the quantum system diverges on the long time scale. 
The model predicts that the energy of a particle with mass $m$ increases linearly in time
with a rate
$
\xi = 3\hbar^2 m \lambda/(4 r^2_C m^2_0).
$
As will become clear by the following analysis,
the reason for such an energy increase is precisely due to the absence
of dissipation within the model.  
The noise acts like an infinite temperature background, steadily increasing
the energy of the system. 

\section{Dissipative extension of the CSL model}\label{sec:deo}

\subsection{Definition of the model via a non-linear stochastic differential equation}

Now that we have clarified the problem of the CSL model we want to work out,
as well as the features that must be preserved,
we are in the position to formulate a new, dissipative CSL model.
As for the original model, the most compact way to do so, is to
define a proper stochastic differential equation.
Specifically, we consider the following non-linear stochastic differential equation:
\begin{eqnarray}\label{eq:sdecsld}
&&\mathd \ket{\varphi_t} = \left[-\frac{i}{\hbar}\hat{H} \mathd t + \frac{\sqrt{\gamma}}{m_0} \int \mathd {\bf y} [\hat{\mathbb{L}}({\bf y})-r_t({\bf y})] \mathd W_t({\bf y}) \right. \\
&&\left.- \frac{\gamma}{2 m^2_0} \int \mathd {\bf y}  [[\hat{\mathbb{L}}^{\dag}({\bf y})\hat{\mathbb{L}}({\bf y})+r^2_t({\bf y})-2r_t({\bf y})\hat{\mathbb{L}}({\bf y})] \mathd t \right] \ket{\varphi_t}, \nonumber
\end{eqnarray}
with $r_t({\bf y}) \equiv \bra{\varphi_t}(\hat{\mathbb{L}}^{\dag}({\bf y})+\hat{\mathbb{L}}({\bf y}))\ket{\varphi_t}/2$ and
\begin{eqnarray}
\hat{\mathbb{L}}({\bf y}) &\equiv& \sum_j \frac{m_j}{(1+k_j)^3} \int  \frac{\mathd {\bf x}}{(\sqrt{2 \pi} r_C )^3} e^{-\frac{|{\bf y}-{\bf x}|^2}{2 r^2_C(1+k_j)^2}} \nonumber\\
&\times& \hat{\psi}_j^{\dag}({\bf x}) \hat{\psi}_j\left(\frac{1-k_j}{1+k_j}{\bf x}+\frac{2 k_j}{1+k_j}{\bf y}\right),\label{eq:ly}
\end{eqnarray}
where
\begin{equation}\label{eq:kj}
k_j \equiv \frac{\hbar}{2 m_j v_{\eta}r_C}.
\end{equation}
The inclusion of dissipation calls for the introduction
of a new parameter, $v_{\eta}$, with the dimension
of a velocity. Crucially, this parameter is related to the temperature of the noise field,
as it will be shown later (see Eq.(\ref{eq:T})), where the numerical choice of $v_{\eta}$ will be also discussed.
The structure of the stochastic differential equation (\ref{eq:sdecsld}) generalizes that of Eq.(\ref{eq:sdecsl})
to the case of non self-adjoint operators \cite{Barchielli2009,Bassi2013b}.
Indeed, for $v_{\eta}\rightarrow \infty$, so that $k_j \rightarrow 0$, one recovers the original CSL model.

The physical meaning of the operator $\hat{\mathbb{L}}({\bf y})$ in Eq.(\ref{eq:ly}) is better understood
by taking into account also its momentum representation. One has
\begin{eqnarray}
\hat{\mathbb{L}}({\bf y}) &=& \sum_j \frac{m_j}{(2 \pi \hbar)^3} 
\int \mathd {\bf P} \mathd {\bf Q}\, \hat{a}^{\dag}_j({\bf P}+{\bf Q}) \, e^{-\frac{i}{\hbar} {\bf Q} \cdot {\bf y}} \nonumber\\
&&\times  \exp\left(-\frac{r^2_C}{2 \hbar^2}\left|(1+k_{j}){\bf Q}+ 2 k_{j} {\bf P}\right|^2\right)
\hat{a}_j({\bf P}), \label{eq:lymom}
\end{eqnarray}
where $\hat{a}^{\dag}_j({\bf P})$ and $\hat{a}_j({\bf P})$ are, respectively, the creation
and annihilation operator of a particle of the type $j$ with momentum ${\bf P}$.
By Eqs.(\ref{eq:ly}) and (\ref{eq:lymom}), we see that the action of the 
collapse noise can be compared to that of an external potential which depends not only on the position, but also on momentum of the system,
thus inducing dissipation.
In particular, since the exchanged momentum $Q_i$ in the spatial direction $i= x, y, z$ has a gaussian distribution peaked around $- 2P_i k_j/(1+k_j)$,
the action of the noise will suppress high momenta,
so that the mean kinetic energy of the system, as well as the mean momentum, is subject to relaxation.
This is explicitly shown in Sec. \ref{sec:era}.
Indeed, contrary to any external field, the collapse noise induces an anti-hermitian coupling with matter,
which is necessary in order to induce localization. In addition, the introduction of dissipation
also leads to an hermitian contribution to the coupling, see Sec. \ref{sec:haa}.

\subsection{Linear stochastic differential equation}

In order to study the solutions and properties of Eq.(\ref{eq:sdecsld}), it is
often convenient to deal with an equivalent linear stochastic differential equation \cite{Bassi2003,Bassi2005}.
Here, we briefly sketch the standard procedure which provides
the linear stochastic differential equation associated with the non-linear one in the form given by Eq.(\ref{eq:sdecsld}).
For a complete treatment, the reader is referred to \cite{Barchielli2009}.

Consider a non-linear stochastic differential equation as Eq.(\ref{eq:sdecsld}).
Recall that $W_t({\bf x})$ denotes an ensemble of independent Wiener processes
defined on a common probability space $(\Omega, \mathcal{F}, \mathbbm{P})$.
Let $B_t(\bf x)$ be the ensemble of processes given by
\begin{equation}
B_t({\bf x}) = W_t({\bf x}) + 2 \int_0^t \mathd s \, r_s({\bf x}),
\end{equation} 
where $r_s({\bf x})$ has been defined after Eq.(\ref{eq:sdecsld}).
Now, by means of the Girsanov theorem one can define a new probability $\mathbbm{Q}$ on 
 $(\Omega, \mathcal{F})$ such that $B_t({\bf x})$ is an ensemble of Wiener processes under $\mathbbm{Q}$ \cite{Barchielli2009}.
 In addition, one can define a random vector $\ket{\psi_t}$ such that
 \begin{equation}\label{eq:norm}
\ket{\varphi_t} = \frac{\ket{{\psi}_t}}{\|{\psi}_t \|},
 \end{equation}
where $\ket{\varphi_t}$ satisfies Eq.(\ref{eq:sdecsld}), while $\ket{{\psi}_t}$ satisfies \cite{Barchielli2009}
\begin{eqnarray}
\mathd \ket{{\psi}_t} &=& \left[-\frac{i}{\hbar}\hat{H}\mathd t-\frac{\gamma}{2 m^2_0}\int \mathd {\bf y}\hat{\mathbb{L}}^{\dag}({\bf y})\hat{\mathbb{L}}({\bf y}) \mathd t \right.  \nonumber\\
& &\left.+ \frac{\sqrt{\gamma}}{m_0} \int \mathd {\bf y} \hat{\mathbb{L}}({\bf y}) \mathd B_t({\bf y}) \right] \ket{{\psi}_t}.
\end{eqnarray}
This is the linear stochastic differential equation associated with the dissipative CSL model.

\subsection{Hermitian and anti-hermitian coupling of the noise field}\label{sec:haa}

In this paragraph, we show that Eq.(\ref{eq:sdecsld}) can be written in a way such that
the coupling between the classical collapse noise and quantum matter is made explicit.
In particular, the introduction of dissipation within the CSL model
leads to an hermitian contribution to the coupling of the collapse noise with quantum matter,
in addition to a contribution in the usual (for collapse models) anti-hermitian form \cite{Bassi2013,Adler2014}.

It is here convenient to use the Stratonovich formalism \cite{Arnold1971,Bassi2003}
and to consider the decomposition of $\hat{\mathbb{L}}({\bf y})$ into its hermitian and anti-hermitian part,
$\hat{\mathbb{L}}({\bf y}) = \hat{\mathbb{L}}^{(a)}({\bf y}) + i \hat{\mathbb{L}}^{(b)}({\bf y})$, with $\hat{\mathbb{L}}^{(a)}({\bf y})$
and $\hat{\mathbb{L}}^{(b)}({\bf y})$ self-adjoint operators.
Thus,
the non-linear stochastic equation given by Eq.(\ref{eq:sdecsld}) leads to
\begin{eqnarray}
\frac{\mathd \ket{\varphi_t}}{\mathd t} &=& \left[-\frac{i}{\hbar}\left(\hat{H}- \frac{\sqrt{\gamma} \hbar}{m_0} \int \mathd {\bf y} \hat{\mathbb{L}}^{(b)}({\bf y})  w({\bf y},t) + R \right)\right. \nonumber\\
&&+\left. \frac{\sqrt{\gamma}}{m_0} \int \mathd {\bf y} \hat{\mathbb{L}}^{(a)}({\bf y})  w({\bf y},t) + S \right] \ket{\varphi_t},\label{eq:eq}
\end{eqnarray}
where $w({\bf x}, t)$ is the white-noise field, which can be formally written as $w({\bf x},t) = \mathd W_t({\bf x})/ \mathd t$
and satisfies the relations
\begin{equation}
\mathbb{E}[w({\bf x},t)] = 0 \qquad \mathbb{E}[w({\bf x},t) w({\bf y},t')] = \delta({\bf x} - {\bf y})\delta(t-t'),
\end{equation}
$\mathbb{E}$ being the stochastic average under the reference probability $\mathbbm{P}$.
Moreover, $R$ is an hermitian contribution coming from the passage to the Stratonovich formalism and reads
\begin{eqnarray}
R &=& - \frac{\gamma \hbar}{m^2_0} \int \mathd {\bf y} \left(  \hat{\mathbb{L}}^{(b)}({\bf y})\hat{\mathbb{L}}^{(a)}({\bf y}) \right.\nonumber\\
 &&\left. -2  \hat{\mathbb{L}}^{(b)}({\bf y})\langle \hat{\mathbb{L}}^{(a)}({\bf y}) \rangle_t 
- \langle \hat{\mathbb{L}}^{(b)}({\bf y}) \hat{\mathbb{L}}^{(a)}({\bf y}) \rangle_t  \right).
\end{eqnarray}
On the other hand, $S$ includes the non-linear contributions preserving the norm of the state vector and is given by
\begin{eqnarray}
S &=& - \frac{\gamma}{m^2_0} \int \mathd {\bf y} \left[\left(  \hat{\mathbb{L}}^{(a)}({\bf y})- \langle \hat{\mathbb{L}}^{(a)}({\bf y}) \rangle_t  \right)^2 + ( \langle \hat{\mathbb{L}}^{(a)}({\bf y}) \rangle_t )^2\right. \nonumber \\
&&\left.- \langle (\hat{\mathbb{L}}^{(a)}({\bf y}))^2 \rangle_t \right]
 -  \frac{\sqrt{\gamma}}{m_0} \int \mathd {\bf y} \langle\hat{\mathbb{L}}^{(a)}({\bf y})\rangle_t w({\bf y},t); 
\end{eqnarray}
compare with Eq.(7.43) in \cite{Bassi2003}.

Equation (\ref{eq:eq}) describes the coupling between the classical field $w({\bf x}, t)$ and quantum matter.
Now, since $\hat{\mathbb{L}}({\bf y})$ is not a self-adjoint operator such a coupling has an hermitian, as well as an anti-hermitian
contribution.
Note that in the original CSL model the collapse noise is coupled with matter only via an anti-hermitian term \cite{Bassi2003,Adler2014}.
To be explicit, 
 Eq.(\ref{eq:lymom}) implies
\begin{eqnarray}
\hat{\mathbb{L}}^{\dag}({\bf y}) & = & \sum_j \frac{m_j}{(2 \pi \hbar)^3} 
\int \mathd {\bf P} \mathd {\bf Q}\, \hat{a}^{\dag}_j({\bf P}+{\bf Q}) \, e^{-\frac{i}{\hbar} {\bf Q} \cdot {\bf y}}\\
&&\times  \exp\left(-\frac{r^2_C}{2 \hbar^2}\left|(1-k_{j}){\bf Q}- 2 k_{j} {\bf P}\right|^2\right)
\hat{a}_j({\bf P}) \nonumber
\end{eqnarray}
and thus
\begin{eqnarray}
\hat{\mathbb{L}}^{(a)}({\bf y})  &=&  \sum_j \frac{ m_j}{(2 \pi \hbar)^3} 
\int \mathd {\bf P} \mathd {\bf Q}\, \hat{a}^{\dag}_j({\bf P}+{\bf Q}) 
\hat{a}_j({\bf P}) \nonumber\\
&&\times \exp\left(-\frac{r^2_C}{2 \hbar^2}\left( |{\bf Q}|^2+k_j^2|{\bf Q}+2{\bf P}|^2  \right)  \right) \nonumber\\
&&\times \cosh \left(\frac{k_j r^2_C}{\hbar^2} {\bf Q}\cdot({\bf Q}+2{\bf P}) \right)
\end{eqnarray}
and 
\begin{eqnarray}
\hat{\mathbb{L}}^{(b)}({\bf y})  &=& - \sum_j \frac{ m_j}{(2 \pi \hbar)^3} 
\int \mathd {\bf P} \mathd {\bf Q}\, \hat{a}^{\dag}_j({\bf P}+{\bf Q}) 
\hat{a}_j({\bf P}) \nonumber\\
&&\times \exp\left(-\frac{r^2_C}{2 \hbar^2}\left( |{\bf Q}|^2+k_j^2|{\bf Q}+2{\bf P}|^2  \right)  \right) \nonumber\\
&&\times \sinh \left(\frac{k_j r^2_C}{\hbar^2} {\bf Q}\cdot({\bf Q}+2{\bf P}) \right).
\end{eqnarray}
Of course, $\hat{\mathbb{L}}^{(b)}({\bf y})=0$ for $k_j =0$.

\section{Energy relaxation}\label{sec:era}

In this section, we investigate the energy behavior of 
a system subjected to the collapse noise in our extended model. 
We deal with the master equation implied by the stochastic differential equation given by Eq.(\ref{eq:sdecsld}).
After presenting the equation in a second-quantization formalism, we describe the corresponding operators
in the case of a fixed number of particles. In particular, by focusing on the one-particle case, we 
show explicitly the exponential relaxation of the energy to a finite value.

\subsection{Master equation for the system's statistical operator}\label{sec:mea}

The non-linear, as well as the linear, stochastic differential equation fully fixes the collapse
model we are defining here. However, one is often interested in the predictions
of the model related with the statistical mean of some physical quantity, 
\begin{equation}\label{eq:o}
O(t) \equiv  \mathbbm{E}[\bra{\varphi_t}\hat{O}\ket{\varphi_t}],
\end{equation}
where, as usual, $\ket{\varphi_t}$ is the stochastic state of the system satisfying Eq.(\ref{eq:sdecsld}).
For this reason, it can be convenient to deal directly with the evolution of the average state
\begin{equation}\label{eq:hat}
\hat{\rho}(t) = \mathbbm{E}[\ket{\varphi_t}\bra{\varphi_t}],
\end{equation}
so that one recovers the usual relation
\begin{equation}
O(t) = \mbox{tr}\left\{\hat{O} \hat{\rho}(t)\right\}.
\end{equation}
By using the It\^{o} calculus, it is easy to see that Eq.(\ref{eq:sdecsld})
implies the following master equation:
\begin{eqnarray}
\frac{\mathd}{\mathd t}\hat{\rho}(t) 
&=&   - \frac{i}{\hbar}\left[\widehat{H} \,,\, \hat{\rho}(t)\right] + 
 \frac{\gamma}{m^2_0} \int \mathd {\bf y}  \left[\hat{\mathbb{L}}({\bf y})\hat{\rho}(t)\hat{\mathbb{L}}^{\dag}({\bf y}) \right. \nonumber\\
 &&\left. - \frac{1}{2}\left\{\hat{\mathbb{L}}^{\dag}({\bf y}) \hat{\mathbb{L}}({\bf y}), \label{eq:mecsldiss}
 \hat{\rho}(t)\right\} \right].
\end{eqnarray}
This is a Lindblad master equation \cite{Lindblad1976,Gorini1976,Breuer2002}, indicating that we are in the presence of a time-homogeneous Markovian dynamics.
The Lindblad operators are the same operators as those appearing in the stochastic differential equation defining the model, see Eq.(\ref{eq:ly}) or Eq.(\ref{eq:lymom}).

The expressions of the Lindblad operators $\hat{\mathbb{L}}({\bf y})$ restricted to a sector of the Fock space
with a fixed number of particles is easily obtained as follows.
Let us assume for simplicity that we have $N$ particles of the same type and mass $m$.
The corresponding restriction of $\hat{\mathbb{L}}({\bf y})$
reads
\begin{eqnarray}
\hat{\mathbb{L}}({\bf y}) &=& \frac{m}{(2 \pi \hbar)^3} \sum^{N}_{\alpha =1} 
\int \mathd {\bf Q}\, e^{\frac{i}{\hbar} {\bf Q} \cdot (\hat{{\bf x}}_{\alpha}- {\bf y})} \nonumber\\
&& \times \exp\left(-\frac{r^2_C}{2 \hbar^2}\left|(1+k){\bf Q}+ 2 k \hat{{\bf P}}_{\alpha}\right|^2\right),
 \label{eq:lybb}
\end{eqnarray}
where $\hat{{\bf x}}_{\alpha}$ and $\hat{{\bf P}}_{\alpha}$ are, respectively, the position
and momentum operator of the $\alpha$-th particle and $k$ is the constant
given by Eq.(\ref{eq:kj}) with $m_j = m$.
Indeed,
consider the Hilbert space $L^{2}(\mathbb{R}^3)$ and the corresponding Fock space $\mathcal{F}(L^{2}(\mathbb{R}^3)) = \mathbb{C} + \sum^{\infty}_{n=1} L^{2}(\mathbb{R}^3)^{\tilde{\otimes}N}$,
where $L^{2}(\mathbb{R}^3)^{\tilde{\otimes}N}$ denotes the symmetric or antisymmetric part of the tensor product $L^{2}(\mathbb{R}^3) \otimes \cdots \otimes L^{2}(\mathbb{R}^3)$, $N$ times.
Now consider the operator on $\mathcal{F}(L^{2}(\mathbb{R}^3))$ given by \cite{Schwabl2008}
\begin{equation}\label{eq:c}
 \hat{\mathbb{A}} = \int \mathd {\bf P} \mathd {\bf P'} \hat{a}^{\dag}({\bf P}') \bra{{\bf P}'} \hat{A}^{(1)}(\hat{{\bf x}}, \hat{\bf P}) \ket{{\bf P}} \hat{a}({\bf P}),
\end{equation}
where $\hat{A}^{(1)}(\hat{{\bf x}}, \hat{\bf P})$  is a single-particle operator on $L^{2}(\mathbb{R}^3)$, with $\hat{{\bf x}}$ and $\hat{\bf P}$, respectively, position and momentum operators on $L^{2}(\mathbb{R}^3)$. 
Hence, the restriction of $\mathbb{A}$ on the $N$-particle sector of the Fock space
reads
\begin{equation}\label{eq:un}
\hat{\mathbb{A}} = \sum^{N}_{\alpha=1} \hat{A}^{(1)}(\hat{{\bf x}}_\alpha, \hat{\bf P}_\alpha),
\end{equation}
$\hat{{\bf x}}_\alpha$ and $\hat{\bf P}_\alpha$ being the position and momentum operator of the $\alpha$-th particle. 
The relation between Eq.(\ref{eq:c}) and Eq.(\ref{eq:un}) is indeed the same as that between
Eq.(\ref{eq:lymom}) and Eq.(\ref{eq:lybb}).

If we further restrict to the case of a single free particle with mass $m$,
we end up with the following
master equation for the one-particle average state $\hat{\rho}^{(1)}$
\begin{eqnarray}
\frac{\mathd}{\mathd t}\hat{\rho}^{(1)}(t) 
&=&   - \frac{i}{\hbar}\left[\frac{{\hat{\bf P}}^2}{2m} \,,\, \hat{\rho}^{(1)}(t)\right] +  \frac{\gamma m^2}{(2 \pi \hbar)^3 m^2_0} \nonumber\\
&&\left(
\int  \mathd {\bf \mom} \, e^{\frac{i}{\hbar} {\bf \mom} \cdot \hat{{\bf x}}} 
L({\bf Q}, \hat{\bf P}) \hat{\rho}^{(1)}(t) 
L({\bf Q}, \hat{\bf P})e^{-\frac{i}{\hbar} {\bf \mom} \cdot \hat{{\bf x}}}\right. \nonumber\\
&&\left. -\frac{1}{2}\left\{L^2({\bf Q}, \hat{\bf P}) , \hat{\rho}^{(1)}(t)\right\}\right),\label{eq:mecsldiss2}
\end{eqnarray}
with
\begin{equation}\label{eq:lqp}
L({\bf Q}, \hat{\bf P}) = e^{-\frac{\rc^2}{2\hbar^2} \left|(1+k) {\bf \mom}+ 2 k \hat{{\bf P}}\right|^2}.
\end{equation}
Let us note that the inclusion of dissipation in the CSL model preserves the
invariance under translations of the system's evolution, but it breaks the invariance under boosts,
as directly seen by the master equation (\ref{eq:mecsldiss2}) \cite{Holevo1993}.
Nevertheless, the characterization of the overall dynamics by means of
a proper first-principle underlying theory,
which involves both the sources of the collapse noise and the quantum systems affected by it,
should allow to recover a fully covariant description; see also the discussion in the next paragraph. 

\subsection{Evolution equation for the average kinetic energy and noise temperature}\label{sec:eef}

The master equation for the system's statistical operator provides us with
the evolution equation of the mean kinetic energy 
of the one-particle system,
\begin{equation}
H(t) = \text{tr}\left\{\hat{{\bf P}}^2/(2m) \hat{\rho}^{(1)}(t)\right\}.
\end{equation}
Exploiting the translation covariance of Eq.(\ref{eq:mecsldiss2}) \cite{Holevo1993,Vacchini2009}
one gets
\begin{eqnarray}
\frac{\mathd}{\mathd t} H(t) &=& 
 \frac{\gamma m}{2(2 \pi \hbar)^3 m^2_0} \int \mathd {\bf \mom}
\text{tr}\left\{e^{-\frac{\rc^2}{\hbar^2} \left|(1+k){\bf \mom}+2 k \hat{{\bf P}}\right|^2}  \right.\nonumber \\
&&\left.\times\left( |{\bf Q}|^2 +2 \hat{{\bf P}} \cdot {\bf Q} \right)\hat{\rho}^{(1)}(t)\right\}, \nonumber\\
&=& \frac{3 \hbar^2 \lambda m}{4(1+k)^5 r^2_C m^2_0}- \frac{4 k \lambda m^2}{(1+k)^5 m^2_0} H(t). \label{eq:hht}
\end{eqnarray}
This equation is solved by
\begin{equation} \label{eq:ht}
H(t) = e^{- \chi t}\left(H(0)-H_{\text{as}} \right) + H_{\text{as}},
\end{equation}
with relaxation rate 
\begin{equation}
\chi = \frac{4 k \lambda m^2}{(1+k)^5 m^2_0}
\end{equation}
and asymptotic kinetic energy 
\begin{equation}
H_{\text{as}} =\frac{3 \hbar^2}{16 k m r^2_C}.
\end{equation}
As expected, now we do have dissipation. The mean energy of the system can
decrease as a consequence of the action of the noise.
Moreover, even if the energy grows, there is an upper threshold value above which it cannot increase. 
The long-time energy divergence is now avoided \cite{foot2}.
This is precisely the result
we wanted and the most natural way to interpret it is to
say that the collapse noise has a finite temperature toward which
the system thermalizes \cite{Bassi2005}. Explicitly,
$H_{\text{as}}$ corresponds to a noise temperature
\begin{equation}\label{eq:T}
T = \frac{\hbar v_{\eta}}{4 k_B r_C},
\end{equation}
where we used Eq.(\ref{eq:kj}) and $k_B$
is the Boltzmann constant.
The original CSL model is recovered in the limit $T \rightarrow \infty$: the noise acts like an infinite temperature background, which 
explains the energy divergence.

The temperature of the noise in Eq.(\ref{eq:T}) does not depend on the mass of the system, which is 
a very important feature of our model. In addition, the 
state of the system actually equilibrates to the canonical Gibbs state, see Appendix \ref{sec:adc}.
These hallmarks of the evolution induced by Eq.(\ref{eq:sdecsld}) depend substantially
on the choice of the operators $\hat{\mathbb{L}}({\bf y}) $ in Eq.(\ref{eq:ly}).
It is an open question to identify the entire class of operators satisfying these natural requests.
In Appendix \ref{sec:adc}, we take into account a physically motivated alternative to the choice made
in Eq.(\ref{eq:ly}), showing how the relaxation dynamics
of the resulting collapse model is very similar to that presented here and,
in particular, the noise temperature is still given by Eq.(\ref{eq:T}).
The exponential relaxation of the energy $H(t)$ in Eq.(\ref{eq:ht}) is 
the same as that in the dissipative Ghirardi-Rimini-Weber (GRW) model recently introduced in \cite{Smirne2014}.
This is not surprising, since, as for the case without dissipation, the extended GRW and CSL
models share the same one-particle master equation.

If we think that the collapse model fixed by Eq.(\ref{eq:sdecsld}) describes 
in a phenomenological way 
the action of a real physical field filling space, it is now clear how the principle
of energy conservation can be reestablished. The energy gained
or lost by the system can be ascribed to an energy exchange with the noise field, as the latter can be influenced back by the presence 
of the system.
An explicit characterization of this process
requires an underlying theory, which has to guarantee the classical nature of the noise field, with its own (non-quantum)
equations of motion, in order to provide a proper objective collapse of the wavefunction \cite{Bassi2003,Bassi2013,Adler2014}.
In addition,
one can already say that a collapse noise with typical cosmological features would correspond to
a low-temperature noise  \cite{Bassi2010,Bassi2013}, at most of the order of few Kelvins. 
By Eq.(\ref{eq:T}), we see that the noise temperature $T$ is in one-to-one
correspondence with the new parameter $v_{\eta}$. For instance, $v_{\eta} = 10^5 \text{m}/\text{s}$ 
(i.e. $k \approx 3 \times 10^{-6}$ for a nucleon)
gives $T \approx 1 K$. 
Hence, more than the specific value of the noise temperature,
the important thing is that even in the presence of a low-temperature noise 
the resulting collapse model can be introduced in a consistent way; see also the discussion in the next section.
It is worth noting that it is not always
possible to properly modify a given collapse model to include dissipation
via the action of a low-temperature noise. In \cite{Bahrami2014b}
we discuss how such a modification is not feasible for the Di{\'o}si-Penrose model~\cite{Diosi1987}, 
in which gravity is involved to provide a phenomenological description of the wavefunction collapse.

In our model,
every system with a temperature higher than about $1 \text{K}$ is cooled by the action of the collapse noise. 
Thus, we are led to reject 
the bounds on the collapse rate $\lambda$ relying on a balance
between the system's heating due to the action of the noise in the original CSL model and the cooling
due to, e.g., the Universe expansion or the energy radiation. 
This is the case for the heating of the protons constituting the intergalactic medium
or for the energy accumulation in interstellar dust grains \cite{Adler2007}.
Note how, in particular, the heating of the IGM provides the second strongest bound to date on the localization rate $\lambda$ \cite{Adler2009}.
Even more, we expect that cosmological
data will put strong bounds on the dissipation parameter $k$ (equivalently, on $v_{\eta}$).
The modified long-time behavior of the energy predicted 
by our model will have to be compared with the constraints coming from
such data. Some preliminary results have been obtained for the non-dissipative CSL model \cite{Adler2007,Lochan2012}. 
Dissipative effects are expected to play an important role also in the  experimental investigation of collapse models via optomechanical systems \cite{Bahrami2014,Nimmrichter2014},
where proper signatures could be visible in the density noise spectrum of the mechanical oscillator, 
or via the spontaneous photon emission from electrically charged particles \cite{Adler2013b,Donadi2014},
as the latter is registered over a period of several years.
In both situations, we expect that dedicated experiments 
should allow to restrict the possible values of $k$; of course,
also in relation with the other parameters of the model.

\section{Macroscopic objects: localization and amplification mechanism}\label{sec:mol}

As recalled in Sec. \ref{sec:tcm},
any proper collapse model is characterized by the amplification mechanism.
The localizing action of the collapse noise has
to increase with size of the system, which guarantees a
consistent description of microscopic and macroscopic systems within a unique theoretical framework.
Here, we show that the amplification mechanism holds in our extended
model, at least as long as one deals with a macroscopic rigid body.
The description of more complex systems, where the internal dynamics
becomes involved, calls for a more detailed specification of the system's evolution \cite{Smirne2014}.
We stress that the following considerations are valid also in the case of a low temperature noise. 
As anticipated, even for a noise temperature $T \approx 1 K$
we have effective localization and amplification mechanisms, so that the noise 
actually induces a classical behavior of the center of mass of macroscopic objects.

Finally, the different role of the momentum-dependent localization operators of our model
in, respectively, the energy relaxation and the wave-function localization is clarified
by comparing the time scales of the two phenomena.

\subsection{Localization of the center of mass of a rigid body and amplification mechanism}

Consider a macroscopic object made up of $N$ particles of equal mass $m$.
One can neglect the contributions due to the electrons and consider the mass of the proton $m_{\text{P}}$
equal to the mass of the neutron $m_{\text{N}}$, i.e. $m_{\text{P}} \approx m_{\text{N}} \equiv m$.
In addition, we deal with a rigid body, which
allows us to decouple the evolution of the center of mass from that of the relative coordinates~\cite{Smirne2014}.
Let $\hat{{\bf x}}_{\text{CM}} = \sum_j \hat{{\bf x}}_j/N$ be the position operator of the center of mass,
while the relative coordinates $\hat{{\bf r}}_j$, $j=1, \ldots, N-1$, are fixed by $\hat{{\bf x}}_j = \hat{{\bf x}}_{\text{CM}}+ \sum_{j'} \Lambda_{j j'} \hat{{\bf r}}_{j'}$,
for a suitable matrix with elements $\Lambda_{j j'}$.
We neglect the possible rotations of the system: this greatly simplifies
the description, without affecting the physical meaning of the results \cite{Ghirardi1990}.
By virtue of the rigid-body approximation, the relative coordinates are fixed
and the center-of-mass momentum $\hat{{\bf P}}_{\text{CM}}$ is simply proportional to each individual momentum $\hat{{\bf P}}_j$,
according to $ \hat{{\bf P}}_j\approx \hat{{\bf P}}_{\text{CM}}/N$, for each $j$.
Moreover, consider a total Hamiltonian $\hat{H} = \hat{H}_{\text{CM}}+ \hat{H}_r$,
given by the sum of two terms associated with, respectively, the center of mass and the relative degrees of freedom.
Thus, the state of center of mass $\ket{\varphi^{(\text{CM})}_t}$ satisfies a stochastic differential equation
with the same form as Eq.(\ref{eq:sdecsld}), where $\hat{H}$ is replaced by $\hat{H}_{\text{CM}}$ and $\hat{\mathbb{L}}({\bf y})$ is replaced by  [compare with Eq.(\ref{eq:lybb})]
\begin{eqnarray}\label{eq:sss}
\hat{\mathbb{L}}^{(\text{CM})}({\bf y}) &=& \frac{m}{(2 \pi \hbar)^3} \int \mathd {\bf Q} \mathcal{F}_r({\bf Q})  e^{\frac{i}{\hbar}  {\bf Q} \cdot (\hat{{\bf x}}_{\text{CM}}-{\bf y})} \\
&&\times  \exp\left(-\frac{r^2_C}{2 \hbar^2}\left|(1+k ){\bf Q}+ 2 k \hat{{\bf P}}_{\text{CM}}/N\right|^2\right), \nonumber
\end{eqnarray}
$\hat{{\bf P}}_{\text{CM}}$ being the center-of-mass momentum operator and
we introduced the function 
\begin{equation}
\mathcal{F}_r({\bf Q}) = \sum_j \exp\left(\frac{i}{\hbar} \sum_{j'}  \Lambda_{j j'} {\bf Q} \cdot {\bf r}_{j'}\right),
\end{equation}
where ${\bf r}_j$ is the fixed $j$-th relative coordinate of the rigid body. The factor $\mathcal{F}_r({\bf Q})$
conveys the influence of the internal structure on the evolution of the center of mass
and it is due to the indistinguishability of particles: it is also present
in the original CSL model \cite{Ghirardi1990}, but not in the GRW models \cite{Ghirardi1986,Smirne2014}.
This factor determines the specific features of the amplification mechanism within the CSL model. 

Hence, we are interested in the action of the operator $\hat{\mathbb{L}}^{(\text{CM})}({\bf y})$ 
on a generic state $\ket{\varphi^{(\text{CM})}}$ of the center of mass and, in particular, we focus
on the changes (if any) in the localization process due to dissipation.
Using the notation$\scalar{{\bf x}}{\varphi^{(\text{CM})}} = \varphi ({\bf x})$
and $\scalar{{\bf P}}{\varphi^{(\text{CM})}} = \tilde{\varphi} ({\bf P})$,
we have the (non-normalized) wave function
\begin{eqnarray}
\phi({\bf x}) &\equiv& \bra{{\bf x}}\hat{\mathbb{L}}^{(\text{CM})}({\bf y}) \ket{\varphi^{(\text{CM})}}\\
&= &
\frac{m}{(2 \pi \hbar)^{9/2}} \int \mathd {\bf Q} \mathd {\bf P} \mathcal{F}_r({\bf Q}) e^{\frac{i}{\hbar}  {\bf Q} \cdot ({\bf x}-{\bf y})}e^{\frac{i}{\hbar}  {\bf P} \cdot {\bf x}} \nonumber\\
&&\times 
 \exp\left(-\frac{r^2_C}{2 \hbar^2}\left|(1+k ){\bf Q}+ 2 k \frac{{\bf P}}{N}\right|^2\right) \tilde{\varphi}({\bf P}). \nonumber
\end{eqnarray}
Now, we use the continuum limit, so that
\begin{equation}
\mathcal{F}_r({\bf Q}) =\int \mathd {\bf z} D({\bf z}) e^{\frac {i}{\hbar} {\bf Q} \cdot {\bf z}},
\end{equation}
$D({\bf z})$ being the macroscopic density of particles, and we obtain
\begin{eqnarray}
\phi({\bf x}) &=& \frac{m}{(2 \pi \hbar)^{3/2}(\sqrt{2 \pi}r_C (1+k)^3)} \int \mathd {\bf z} \mathd {\bf P} D({\bf z}) e^{\frac{i}{\hbar}  {\bf P} \cdot {\bf x}} \nonumber\\
&& \exp\left(-\frac{|{\bf x} - {\bf y} + {\bf z}|^2}{2 r^2_C(1+k)^2}\right)e^{-\frac{2 i k}{(1+k)\hbar N} {\bf P} \cdot ({\bf x} - {\bf y} + {\bf z})} \tilde{\varphi}({\bf P}). \nonumber
\end{eqnarray}
Since 
$$
\varphi({\bf x}) = \int \mathd {\bf P} \frac{e^{\frac{i}{\hbar} {\bf P} \cdot {\bf x}}}{(2 \pi \hbar)^{3/2}} \tilde{\varphi}({\bf P}),
$$
one gets
\begin{eqnarray}
\phi({\bf x}) &=& \frac{m}{(2 \pi r^2_C)^{3/2}} \int \mathd {\bf z} D({\bf z}) \frac{\exp\left(-\frac{|{\bf x} - {\bf y} + {\bf z}|^2}{2 r^2_C(1+k)^2}\right)}{(1+k)^3} \nonumber\\
&& \times \varphi\left({\bf x} - \frac{2 k}{(1+k)N} ({\bf x} - {\bf y} + {\bf z})\right). \label{eq:expv}
\end{eqnarray}
We now assume that this macroscopic density $D({\bf y})$ does not vary significantly on the length-scale
fixed by $r_C(1+k)$, so that
the exponential in Eq.(\ref{eq:expv}) varies as a function of ${\bf z}$ much faster than the other terms within the integral \cite{foot4}. 
Thus, we can make the substitution
\begin{equation}
 \frac{1}{(2 \pi r^2_C(1+k)^2)^{3/2}}
\exp\left(-\frac{|{\bf x} - {\bf y} + {\bf z}|^2}{2 r^2_C(1+k)^2}\right) \rightarrow \delta^3\left({\bf x} - {\bf y} + {\bf z}\right)
\end{equation}
in Eq.(\ref{eq:expv}), getting
\begin{equation}\label{eq:agia}
\phi({\bf x}) \approx m D({\bf y}-{\bf x}) \varphi({\bf x}).
\end{equation}
This clearly shows how the effects on the localization process due to the presence of the momentum operator in $\hat{\mathbb{L}}^{(\text{CM})}({\bf y})$
can be safely neglected, thus guarantying the convergence toward well localized states.
The localization of the wavefunction, as, e.g., represented in Fig.\ref{fig:1}, is basically not modified
by the introduction of dissipation in the model.

One can draw the same conclusion by dealing with the master equation of the average state,  and, in addition,
one recovers Eq.(\ref{eq:G}) as an effective characterization of the scaling of the localization
strength with the size of the system.
The Lindblad master equation for the state of the center of mass is given by [compare with Eq.(\ref{eq:mecsldiss})]
\begin{eqnarray}
\frac{\mathd}{\mathd t}\hat{\rho}^{(\text{CM})}(t) 
&=&   - \frac{i}{\hbar}\left[\widehat{H}_{\text{CM}} \,,\, \hat{\rho}^{(\text{CM})}(t)\right] \nonumber\\
&& + \frac{\gamma}{m^2_0} \int \mathd {\bf y}  \left[\hat{\mathbb{L}}^{(\text{CM})}({\bf y})\hat{\rho}^{(\text{CM})}(t)\hat{\mathbb{L}}^{(\text{CM})\dag}({\bf y})\right. \nonumber\\
&&\left. - \frac{1}{2}\left\{\hat{\mathbb{L}}^{(\text{CM}) \dag}({\bf y}) \hat{\mathbb{L}}^{(\text{CM})}({\bf y}),
\hat{\rho}^{(\text{CM})}\right\} \right].  \label{eq:mecsldisscm}
\end{eqnarray}
By using Eq.(\ref{eq:agia}) and neglecting the free Hamiltonian contribution,
we end up with the equation in the position representation
\begin{equation}
\partial_t \bra{{\bf x}'} \hat{\rho}^{(\text{CM})}(t) \ket{{\bf x}''} = - \Lambda({\bf x'}, {\bf x}'') \bra{{\bf x}'} \hat{\rho}^{(\text{CM})}(t) \ket{\bf x}'',
\end{equation}
where 
\begin{equation}
 \Lambda({\bf x'}, {\bf x}'') \approx \gamma \int \mathd {\bf y}[D^2({\bf y})-D({\bf y}) D({\bf y}+{\bf x}'-{\bf x}'')].
\end{equation}
The same expression was obtained for the original CSL model in \cite{Ghirardi1990}, 
under the so-called sharp-scanning approximation.
In particular, if we consider a rigid body with constant density $D$, we get \cite{Ghirardi1990}
\begin{equation}
\Lambda({\bf x'}, {\bf x}'') = \gamma D n_{\text{out}},
\end{equation}
with $n_{\text{out}}$ the number of particles of the body when its center of mass
is in the position ${\bf x}'$ that are outside the volume
occupied by the body when its center of mass is in ${\bf x}''$.
Indeed, if $n_{\text{out}}$ is equal to the total number of particles (i.e.
there is no overlap between the volumes occupied by the macroscopic rigid body when its center of
mass is in, respectively, ${\bf x}'$
and ${\bf x}''$), one recovers Eq.(\ref{eq:G}), up to an irrelevant constant factor $(4 \pi)^{3/2}$.
The localization rate, which is vey small for microscopic systems increases
with the size of the system proportionally to the square of the number of particles,
which is a direct signature of the action of the noise on indistinguishable particles.

\subsection{Collapse rate versus relaxation rate}

The comparison between the dissipation rate $\chi$ and the localization rate $\Gamma$, see Eq.(\ref{eq:G}),
shows how the two phenomena occur on different time scales:
while the center of mass of a macroscopic system will be quickly localized by the action
of the noise, dissipation can possibly play a role on the system's evolution only on the long time scale.

We take into account the evolution of the center-of-mass
energy of a macroscopic rigid body with $N$ nucleons, $H^{(\text{CM})}(t) = {tr}\left\{\hat{{\bf P}}^2/(2M) \hat{\rho}^{(\text{CM})}(t)\right\}$, where $M = N m_0$
is the total mass.
If we repeat the calculations performed in Sects.\ref{sec:eef},
the master equation~(\ref{eq:mecsldisscm}), at first order in $k$,
leads to an exponential relaxation of the energy
with rate
\begin{equation}\label{eq:chi}
\chi \approx \frac{16 \sqrt{2} k r_C^5 \lambda}{ (2 r^2_C + R^2)^{5/2}},
\end{equation}
where we considered a spherical object with radius radius $R$ and constant density
and we made the following approximation \cite{Bahrami2014}
\begin{eqnarray}
\mathcal{F}_r({\bf Q}) &=& \frac{3 N}{4 \pi R^3} \left(\sin\left(\frac{\mom R_0}{\hbar}\right)-\frac{\mom R_0}{\hbar}\cos\left(\frac{\mom R_0}{\hbar}\right)\right) \nonumber\\
& \approx & e^{- \frac{Q^2 R^2}{2 \hbar^2}}. \nonumber
\end{eqnarray}
The dissipation rate $\chi$ in Eq.(\ref{eq:chi}) is much smaller than the corresponding localization rate,
which, according to Eq.(\ref{eq:G}), is given by 
\begin{equation}
\Gamma = \lambda n^2 \tilde{N} = \lambda \left(\frac{N r^3_C}{4 \pi R^3/3}\right)^2 \frac{4 \pi R^3/3}{r_C^3} = \lambda \frac{N^2 r^3_C}{4 \pi R^3/3}.
\end{equation}
The ratio between the two rates in the case $R \gg r_C$ is 
\begin{equation}
\frac{\Gamma}{\chi} \approx 10^{4} N^2 \left(\frac{R}{r_C}\right)^2.
\end{equation}
If we consider
a reference density $D = 5 \,{\text g}\, {\text cm}^{-3}$, one has $N \approx 10^{25} (R[{\text cm}])^3$,
since $1 \text{g}$ of matter contains approximately an Avogadro's number of nucleons.
Now, set a radius $R = 1 \text{mm}$, so that $N \approx 10^{22}$.
In this case, the localization rate is $\Gamma = 10^{14} \text{s}^{-1}$, while the dissipation rate is $\chi=10^{-41}\text{s}^{-1}$: the noise
localizes the center of mass of the macroscopic body on very short time scales, while
the influence of dissipation can be safely neglected during the whole evolution of the macroscopic system.
Similarly, if $R = r_C = 10^{-5} \text{cm}$, implying $N \approx 10^{10}$,
we get $\chi = 10^{-22} \text{s}^{-1}$, so that dissipation can still be neglected, while in this case $\Gamma \approx 10^{2}\text{s}^{-1}$.
Moreover, one could wonder how this analysis changes if we choose a different one-particle
localization rate $\lambda$. For the value proposed by Adler \cite{Adler2007}, $\lambda = 10^{-9}\text{s}^{-1}$,
we have that dissipation can still be neglected for $R = 1 \text{mm}$, where $\chi = 10^{-33}\text{s}^{-1}$
(and $\Gamma = 10^{22} \text{s}^{-1}$).
Instead, for $R = r_C = 10^{-5} \text{cm}$, we end up with $\chi = 10^{-14}\text{s}^{-1}$,
so that dissipation can play a role on the secular evolution of the system.
However, also in this case the effect of dissipation on the localization of the wavefunction is completely negligible.
Localization occurs on a much shorter time scale than dissipation, $\Gamma = 10^{10} \text{s}^{-1}$, and then the influence 
of the dissipative terms in Eq.(\ref{eq:expv}) can be neglected to study localization, even if it can subsequently
play a role in the long-time behavior of the system.

\section{Conclusions}\label{sec:c}

The main purpose of collapse models is to provide
a unified framework for the description
of microscopic and macroscopic systems, 
thus avoiding an ad-hoc dividing line within the theory,
as well as yielding a dynamical explanation for the collapse of the wavefunction. 
The results of this paper point out that this program can be followed 
taking into account basic physically-motivated demands.

We have included dissipation in the CSL model, which is up to now
the most refined collapse model. This allowed us to remove the divergence
of the energy on the long time scale affecting the original CSL model. 
This divergence
traces back to an infinite temperature of the collapse noise, which is of course
an unrealistic feature of the model. 
The inclusion of dissipation brings along a new parameter, which is strictly related with the finite
temperature of the noise. 
Significantly, even in the presence of a low-temperature noise
the localization and the amplification mechanism are effective, so that the unified
description of microscopic and macroscopic systems is still guaranteed.

A realistic description of the wavefunction collapse can be further developed, for example
by also including a non-white noise \cite{Adler2007b,Adler2008} within the model.
Nevertheless, one should keep in mind that the specific features of 
the collapse noise can be fixed only through a first-principle 
underlying theory, which can clarify the physical origin of the noise \cite{Adler2004,Bassi2013}. 
The development of such an underlying theory is one of the main goals of the research on collapse models and, more in general,
on the theories going beyond quantum mechanics.

\begin{acknowledgments}
The authors acknowledge financial support by the EU
project NANOQUESTFIT, INFN, FRA-2014 by the University of Trieste and
COST (MP1006). They also thank M. Bahrami, L. Di{\'o}si, S. Donadi, L. Ferialdi, and B. Vacchini
for many useful discussions.

\end{acknowledgments}

\appendix

\section{A different choice of the collapse operators and equilibration to the Gibbs state} \label{sec:adc}
In this Appendix, we discuss an alternative and physically motivated choice
for the operators $\hat{\mathbb{L}}({\bf y})$, which differs from that made in Eq.(\ref{eq:ly}) (or, equivalently, in Eq.(\ref{eq:lymom})).
We will see how this different choice
does not affect significantly the features of the relaxation dynamics we are interested in.
We also argue that in both cases the system's state equilibrates to the stationary solution
in the Gibbs form.

Let us consider the following operators:
\begin{eqnarray}\label{eq:lymom2}
\hat{\mathbb{L}}({\bf y}) &=& \sum_j \frac{m_j}{(2 \pi \hbar)^3} 
\int \mathd {\bf P} \mathd {\bf Q}\, \hat{a}^{\dag}_j({\bf P}+{\bf Q}) \, e^{-\frac{i}{\hbar} {\bf Q} \cdot {\bf y}} \\
&& \times \exp\left(-\frac{r^2_C}{2 \hbar^2}\left((1+k_{j}) Q+ 2 k_{j} \frac{{\bf P} \cdot {\bf Q}}{Q} \right)^2\right)
\hat{a}_j({\bf P}),\nonumber 
\end{eqnarray}
where $Q$ denotes the modulus of ${\bf Q}$. Indeed,
such operators still reduce to those of the original CSL model \cite{Ghirardi1990} in the limit $k_j \rightarrow 0$, i.e. $v_{\eta} \rightarrow \infty$.
The difference with respect to the operators in Eq.(\ref{eq:lymom}) is that now the action of the noise no longer damps
the momentum of the system in an isotropic way, but it tends to suppress the momentum
of a particle in the state $\ket{{\bf P}}$ mainly in the direction ${\bf P} / P$.
A significant motivation for collapse operators as in Eq.(\ref{eq:lymom2}) would be their
correspondence with the operators used to describe the collisional decoherence in the presence
of dissipation. In particular, let us consider for simplicity the one-particle case.
Then, the stochastic differential equation (\ref{eq:sdecsld}), along with the operators in Eq.(\ref{eq:lymom2}),
provides us with a master equation for the average one-particle state $\hat{\rho}^{(1)}$ given by Eq.(\ref{eq:mecsldiss2}), where $L({\bf Q}, \hat{\bf P})$
is now
\begin{equation}\label{eq:lqp2}
L({\bf Q}, \hat{\bf P}) = e^{-\frac{\rc^2}{2\hbar^2} \left((1+k)  \mom+ 2 k \frac{\hat{{\bf P}}\cdot {\bf Q}}{Q}\right)^2}.
\end{equation}
This master equation is the same as that describing the collisional dynamics of a test particle 
interacting with a low-density gas in the weak coupling regime \cite{Vacchini2000}, 
once we make a proper identification of the coefficients in the two models.
In particular, the new parameter $v_{\eta}$ corresponds to the most probable velocity $\sqrt{2/(\beta m_g)}$
of the gas particles, which have mass $m_{g}$ and a Maxwell-Boltzmann momentum distribution,
see \cite{Vacchini2007,Smirne2014} for more details.

By Eqs.(\ref{eq:mecsldiss2}) and (\ref{eq:lqp2}), we directly have that the mean kinetic energy of the system evolves
according to
\begin{eqnarray}
\frac{\mathd}{\mathd t} H(t) &=&  \frac{3 \hbar^2 \lambda m}{4 (1+k)^5  r^2_C m^2_0} \\
&&- \frac{4(1-k) k \lambda m^2}{(1+k)^5 m^2_0} H(t) -\frac{16 k^3 r^2_C M \lambda}{5 \hbar^2 (1+k)^5 m^2_0}
\langle P^4 \rangle_t,  \nonumber
\end{eqnarray}
which involves the forth order momentum of the momentum operator and hence does not yield a closed
equation for the mean kinetic energy. However, at first order in $k$
this equation is exactly the same as Eq.(\ref{eq:hht}), so that one recovers an exponential relaxation
to a finite asymptotic energy which corresponds to the noise temperature in Eq.(\ref{eq:T}).

Even more, this expression of the noise temperature is proved exactly by studying the relaxation
to the equilibrium of the system's state. 
By direct substitution, one can easily see that the canonical Gibbs distribution
\begin{equation}\label{eq:sp}
\varrho(\hat{\bf P}) = \left(\frac{\beta}{2 m \pi}\right)^{3/2} e^{-\frac{\beta \hat{|{\bf P}|}^2}{2 m}} 
\end{equation}
is a stationary solution of Eq.(\ref{eq:mecsldiss2}), with $L({\bf Q}, \hat{\bf P})$ as in Eq.(\ref{eq:lqp2}), if $\beta = 1/(k_B T)$ is the inverse temperature
corresponding to Eq.(\ref{eq:T}). 
Relying on the theorems by Spohn \cite{Spohn1976},
we conclude that this stationary solution is unique 
and, for any initial condition, the state $\hat{\rho}^{(1)}(t)$
converges to it. Hence, the temperature in Eq.(\ref{eq:T}) is the temperature toward which the system
thermalizes and is thus identified as the noise temperature. 
The theorems by Spohn apply when the set of Lindblad operators $V = \left\{L_j\right\}_{j = 1 \ldots n}$
of a given Lindblad generator are such that only the multiples of the identity commute
with all the elements of $V$ and that if $L_j$ is contained in $V$, then also $L^{\dag}_j$ is so.
Indeed, these two conditions are satisfied by set of Lindblad operators in Eqs. (\ref{eq:mecsldiss2}) and (\ref{eq:lqp2}).
To be precise, the theorems by Sphon hold 
in the finite dimensional case, so that, to put the above discussion on a firm mathematical ground,
one should confine the whole system in a finite volume (which would also allow to take $\varrho(\hat{\bf P})$ as a proper quantum state)
and include a cut-off on the momenta.
In addition, once the uniqueness of the stationary state is guaranteed, 
the convergence to it for any initial condition
can be alternatively proved by exploiting the contractivity of the relative entropy under completely positive
and trace preserving maps \cite{Vacchini2009}.

The same results can be applied to our dissipative CSL model, 
i.e., if $L({\bf Q}, \hat{\bf P})$ is given by Eq.(\ref{eq:lqp}). 
Beside providing a further proof of the relation in Eq.(\ref{eq:T}),
this shows that
our model predicts a convergence of the average one-particle state of the system 
to the stationary solution in the canonical form.



\begin{thebibliography}{10}
\bibitem{foot2}
Since the average mean kinetic energy is finite, the mean kinetic energy is almost surely finite
on each trajectory.
\end{thebibliography}
\end{document}